\documentclass[prl, preprintnumbers,showpacs,amsmath,amssymb, superscriptaddress,onecolumn,preprint]{revtex4-2}

\usepackage{graphicx}
\usepackage{dcolumn}
\usepackage{bm}
\usepackage{color}
\usepackage[compact]{titlesec}
\usepackage{hyperref}
\usepackage{changes}
\setcitestyle{super}
\makeatletter
\def\blfootnote{\xdef\@thefnmark{}\@footnotetext}
\makeatother
\begin{document}

\title{Large asymmetric anomalous Nernst effect in the antiferromagnet SrIr$_{0.8}$Sn$_{0.2}$O$_3$ }
\author{Dongliang Gong$\,^{\star\ast\dagger}$\blfootnote{$^{\dagger}\,$Present address: Key Laboratory of Applied Superconductivity, Institute of Electrical Engineering, Chinese Academy of Sciences, Beijing 100190, People’s Republic of China}}
\affiliation{Department of Physics and Astronomy, University of Tennessee, Knoxville, Tennessee 37996, USA}
\author{Junyi Yang$\,^{\star\ddagger}$\blfootnote{$^{\ddagger}\,$Present address: Argonne National Laboratory, Illinois, 60439, USA}}
\affiliation{Department of Physics and Astronomy, University of Tennessee, Knoxville, Tennessee 37996, USA}
\author{Shu Zhang}
	\affiliation{Institute for Theoretical Solid State Physics, IFW Dresden, Helmholtzstrasse 20, 01069 Dresden, Germany}
\author{Shashi Pandey}
\affiliation{Department of Physics and Astronomy, University of Tennessee, Knoxville, Tennessee 37996, USA}
\author{Dapeng Cui}
\affiliation{Department of Physics and Astronomy, University of Tennessee, Knoxville, Tennessee 37996, USA}
\author{Jacob P. C. Ruff}
\affiliation{CHESS, Cornell University, Ithaca, New York 14853, USA}
\author{Lukas Horak}
\affiliation{Department of Condensed Matter Physics, Faculty of Mathematics and Physics, Charles University, Ke Karlovu 5, 121 16 Prague 2, Czech Republic}
\author{Evguenia Karapetrova}
\affiliation{Advanced Photon Source, Argonne National Laboratory, Argonne, Illinois, 60439, USA}
\author{Jong-Woo Kim}
\affiliation{Advanced Photon Source, Argonne National Laboratory, Argonne, Illinois, 60439, USA}
\author{Philip J. Ryan}
\affiliation{Advanced Photon Source, Argonne National Laboratory, Argonne, Illinois, 60439, USA}
\author{Lin Hao}
\affiliation{Anhui Key Laboratory of Condensed Matter Physics at Extreme Conditions, High Magnetic Field Laboratory, HFIPS, Anhui, Chinese Academy of Sciences, Hefei 230031, China}
\author{Yang Zhang$\,^{\ast}$}
\affiliation{Department of Physics and Astronomy, University of Tennessee, Knoxville, Tennessee 37996, USA}
\affiliation{Min H. Kao Department of Electrical Engineering and Computer Science, University of Tennessee, Knoxville, Tennessee 37996, USA}
\author{Jian Liu$\,^{\ast}$}
\affiliation{Department of Physics and Astronomy, University of Tennessee, Knoxville, Tennessee 37996, USA}
\blfootnote{$^{\ast}$Corresponding author: dgong@mail.iee.ac.cn, yangzhang@utk.edu, jianliu@utk.edu}
\blfootnote{$^{\star}$These authors contributed equally: Dongliang Gong, Junyi Yang}

\date{\today}

\begin{abstract}
A large anomalous Nernst effect is essential for thermoelectric energy-harvesting in the transverse geometry without external magnetic field. It’s often connected with anomalous Hall effect, especially when electronic Berry curvature is believed to be the driving force. This approach implicitly assumes the same symmetry for the Nernst and Hall coefficients, which is however not necessarily true. Here we report a large anomalous Nernst effect in antiferromagnetic SrIr$_{0.8}$Sn$_{0.2}$O$_3$ that defies the antisymmetric constraint on the anomalous Hall effect imposed by the Onsager reciprocal relation. The observed spontaneous Nernst thermopower quickly reaches the sub-$\mu$V/K level below the N\'{e}el transition around 250 K, which is comparable with many topological antiferromagnetic semimetals and far excels other magnetic oxides. Our analysis indicates that the coexistence of significant symmetric and antisymmetric contributions plays a key role, pointing to the importance of extracting both contributions and a new pathway to enhanced anomalous Nernst effect for transverse thermoelectrics.
\end{abstract}

\maketitle

\section{Introduction}

Thermoelectric effects directly convert heat into electricity, which is crucial for energy harvesting
\cite{snyder2008complex,bell2008cooling,zhu2017compromise,muchler2013topological}. 
The focus is traditionally on the Seebeck effect, where a temperature gradient generates a longitudinal voltage \cite{pei2012band,shi2020advanced}. The Nernst effect, on the other hand, corresponds to the voltage generated in the transverse direction \cite{behnia2016nernst}, which is usually smaller and requires magnetic field but has unique advantages in efficiency due to the orthogonal geometry \cite{mizuguchi2019energy,washwell1970nernst,goldsmid1972measurement,he2021large}. 
To generate the transverse voltage without magnetic field, magnetic materials have been used to enable the anomalous Nernst effect (ANE) by utilizing the spontaneous time-reversal symmetry-breaking field \cite{guin2019zero,sakai2020iron}. Recent attentions have turned to antiferromagnetic topological materials \cite{xu2020finite,ikhlas2017large,you2022anomalous,pan2022giant} due to the possibility of exploiting the electronic Berry curvatures for large ANE \cite{zhou2020giant,ikhlas2017large,pan2022giant,noky2020giant} and incorporating that with the strengths of antiferromagnetic (AFM) spintronics \cite{jungwirth2016antiferromagnetic,baltz2018antiferromagnetic} for efficient transverse thermoelectrics. Examples include Mn$_{3}$Ge \cite{xu2020finite}, Mn$_{3}$Sn \cite{ikhlas2017large}, Mn$_{3}$SnN \cite{you2022anomalous} and YbMnBi$_{2}$ \cite{pan2022giant}.

\begin{figure}[htbp]
\includegraphics[scale=0.8]{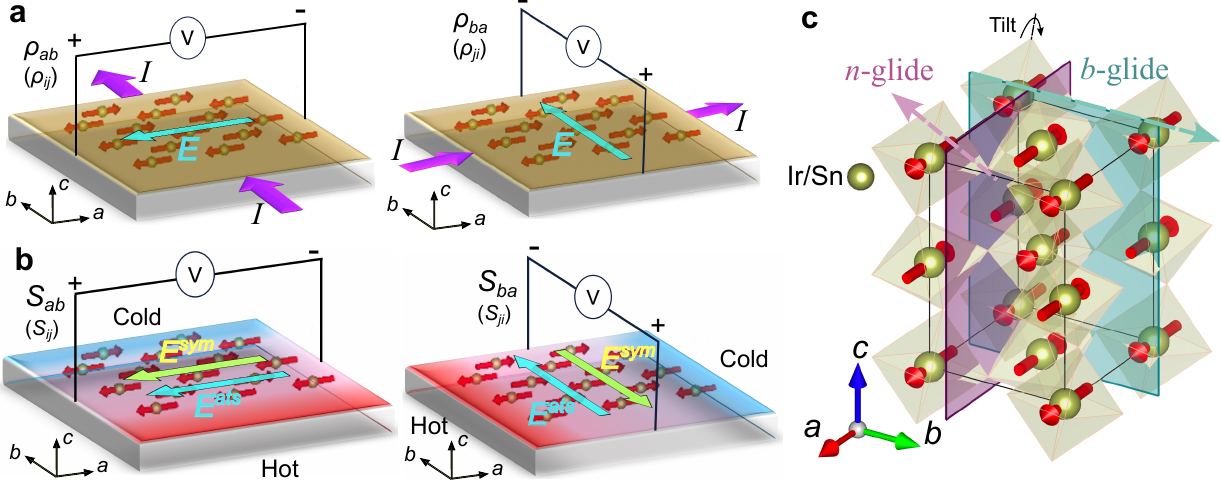}
\caption{\textbf{Illustration of transverse configurations of AHE and ANE measurements on SrIr$_{0.8}$Sn$_{0.2}$O$_3$.} \textbf{a},  The schematic of the two Hall configurations upon exchanging the directions of the applied current (\textit{I}) and the measured voltage (\textit{V}). The general labels of the two directions, $i$ and $j$, are replaced with the crystallographic directions, $a$ and $b$ (Fig.~\ref{figT}{\bf c}), used in the actual measurements. The aqua arrows represent the transverse electric field (\textit{E}). The antisymmetry of the two configurations manifests as a result of the fact that the polarity of $V$ is defined by the coordinate and must be parallel with the polarity of $E$ in one configuration while anti-parallel in the other. The red arrows denote the AFM order of SISO (Fig.~\ref{figT}{\bf c}). \textbf{b}, The schematic of the two Nernst configurations upon exchanging the directions of the temperature gradient and the measured voltage. The red and blue shades represent the hot and cold end, respectively. The aqua and green arrows represent the antisymmetric and symmetric components of the transverse electric field, respectively. \textbf{c}, Crystal structure of SISO. The $n$-glide (orchid) and $b$-glide (cyan) planes are nonsymmorphic operations that protect the semimetal state. The red arrows denote the Ir moments along the $a$-axis in the G-type AFM state which breaks both glide symmetries. }\label{figT}
\end{figure}

A common strategy of finding and investigating such large ANE materials is to look for the anomalous Hall effect (AHE) because Berry curvature of the electronic structure contributes to both effects \cite{xiao2006berry,nagaosa2010anomalous}. Experimentally, AHE is relatively easier to measure as well. However, the general symmetry requirements for ANE and AHE are different. It is known that AHE must comply with the Onsager reciprocal relations regardless of its underlying mechanism \cite{casimir1945onsager,onsager1931reciprocal}, i.e., anomalous Hall resistivity $\rho_{ij}(\textbf{M})=\rho_{ji}(-\textbf{M})$ with $\textbf{M}$ being the magnetic order parameter and $i$ and $j$ representing the two orthogonal directions. In other words, as a time-reversal odd phenomenon, AHE is necessarily antisymmetric upon exchanging the roles of the two axes, i.e., $\rho_{ij}(\textbf{M})=-\rho_{ji}(\textbf{M})$ [Fig.~\ref{figT}{\bf a}]. This requirement is not applicable to ANE \cite{cracknell2016magnetism}, even though it is also time-reversal odd, because the thermoelectric constraint from the Onsager reciprocal relations is imposed between the Nernst effect and the Ettingshausen effect rather than between the different Nernst coefficients.
The generated electric field of ANE may very well have a symmetric component ($E^{\mathrm{sym}}_{ij}(\textbf{M})=E^{\mathrm{sym}}_{ji}(\textbf{M})$) in addition to the antisymmetric one ($E^{\mathrm{ats}}_{ij}(\textbf{M})=-E^{\mathrm{ats}}_{ji}(\textbf{M})$). Their combination implies that the two ANE configurations ($S_{ij}$ and $S_{ji}$) can be highly asymmetric because one can be significantly enhanced at the expense of the other [Fig.~\ref{figT}{\bf b}], which is in sharp contrast to AHE. Therefore, although Berry curvature is antisymmetric \cite{xiao2006berry,nagaosa2010anomalous}, assuming that ANE is antisymmetric as well could miss out on important contributions and new mechanisms to engineer and enhance ANE. Here we report an example of large ANE due to coexistence of both symmetric and antisymmetric contributions observed in AFM Sn-substituted SrIrO$_3$ (SISO). While the AHE of SISO is fully antisymmetric, the presence of a significant symmetric contribution to ANE in addition to an antisymmetric one results in more than one order of magnitude difference in the anomalous Nernst conductivity of the two transverse configurations, giving rise to an ANE thermopower that is well above other reported magnetic oxides and comparable to many magnetic topological metals.

\section*{Results}
Perovskite SrIrO$_3$ is a paramagnetic nonsymmorphic semimetal [Fig.~\ref{figT}{\bf c}], where the so-called $J_{\mathrm{eff}}$ = 1/2 electrons near the Fermi level are at the verge of an instability toward an AFM insulating state due to electron-electron interaction \cite{moon2008dimensionality,zeb2012interplay,carter2012semimetal,nie2015interplay}. The instability can be triggered by 10\% to 30\% isovalent substitution of Ir$^{4+}$ ions with nonmagnetic Sn$^{4+}$ ions [Fig.~\ref{figT}{\bf c}] \cite{cui2016slater,fujioka2018charge}, although the microscopic theory of the Sn substitution is not yet clear.
While bulk SISO is only available in powder/polycrystalline form, single crystalline samples can be obtained by epitaxial stabilization \cite{yang2019epitaxial,negishi2019contrasted}. However, due to the orthorhombicity of SISO, Nernst measurements further require the samples to have a single orthorhombic domain to distinguish the different matrix elements of the thermoelectric tensor, which as we show below is crucial in separating the symmetric and antisymmetric contributions. We thus chose the (001)-oriented TbScO$_3$ [TSO(001)] single crystal substrates to synthesize 20\%-substituted SISO thin films because they both have the same orthorhombic perovskite structure with the \textit{Pbnm} space group \cite{yang2019epitaxial,negishi2019contrasted}. 

\subsection{Single orthorhombic domain structure}
\begin{figure}[htbp]
\includegraphics[scale=0.75]{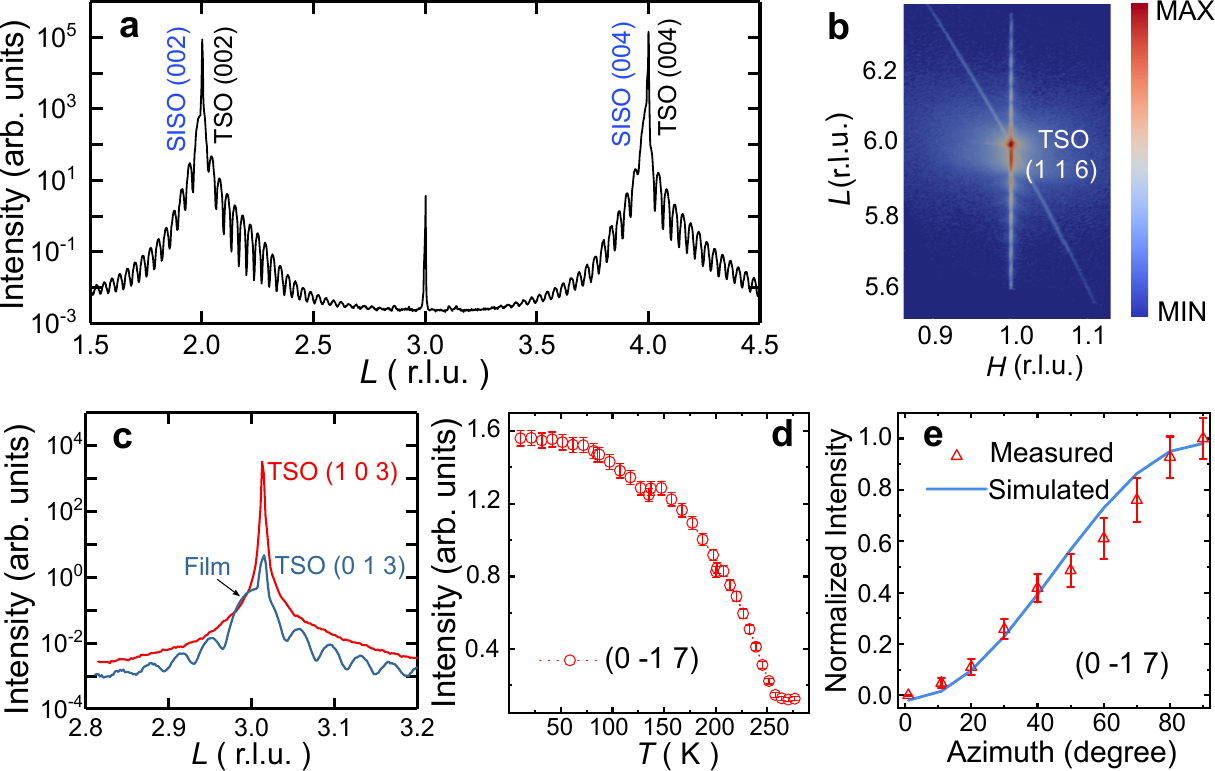}
\caption{\textbf{SrIr$_{0.8}$Sn$_{0.2}$O$_3$ thin film of single orthorhombic crystallographic and magnetic domain.} \textbf{a}, X-ray diffraction (XRD) pattern of Sn-substituted SrIrO$_3$ along [0 0 $L$] direction. The film peak is overlapping with the left side of the TSO substrate peak. The reciprocal lattice unit (r.l.u.) is defined using the TSO substrate. \textbf{b}, Reciprocal space mapping (RSM) around the (1 1 6) reflection. The film peak is partially overlapping with the TSO substrate (1 1 6), and indicates a fully saturated state. \textbf{c}, $L$-scans around the (1 0 3) and (0 1 3) positions of the substrate. The latter is forbidden and hence strongly suppressed. The film peak extinction is however opposite. \textbf{d}, Temperature dependence of the magnetic peak around (0 -1 7). \textbf{e}, Azimuthal dependence of the normalized intensity of the magnetic peak (0 -1 7) at 7 K. The error bars represent the statistical error.}\label{fig1}
\end{figure}

The representative X-ray diffraction results of about 17.4 nm-thick samples (Supplementary Fig.S1) in Fig.~\ref{fig1}{\bf a}\&{\bf b} show high epitaxial quality with pronounced Kiessig fringes in the specular scan and a fully strained state in the reciprocal space mapping. To verify the single orthorhombic domain structure, we measured the IrO$_{6}$ octahedral tilt which is the key distortion for the orthorhombicity [Fig.~\ref{figT}{\bf c}]. Interestingly, as seen in Fig.~\ref{fig1}{\bf c}, no film peak was observed near the (1 0 3) octahedral tilt peak of the substrate. Instead, a clear film peak was observed near the (0 1 3) position of the substrate which is an extinction reflection of the \textit{Pbnm} structure. This observation indicates that the film does have a single orthorhombic domain but with its $a$- and $b$-axes rotated by 90$^o$ from that of the substrate. Hereafter we refer $a$, $b$, and $c$ to the axes of the SISO \textit{Pbnm} cell. 
The single orthorhombic domain facilitates the determination of the spin axis of the AFM structure. Figure~\ref{fig1}{\bf d} shows the temperature dependence of the (0 $-1$ 7) magnetic reflection (Supplementary Fig.S2{\bf a}\&{\bf b}), which was measured by resonant X-ray magnetic scattering (See method) and is characteristic of the G-type AFM structure in \textit{Pbnm} perovskite \cite{yang2019epitaxial}. The N\'{e}el transition was clearly observed around 250 K, which is slightly lower than that bulk powder samples but higher than the film of the same composition on SrTiO$_3$, likely due to the much smaller compressive strain from TSO(001) \cite{lupascu2014tuning,yang2020strain,negishi2019contrasted}. The azimuthal dependence of the (0 $-1$ 7) magnetic peak shown in Fig.~\ref{fig1}{\bf e} indicates that the intensity vanishes at 0 degree where the $a$-axis is perpendicular to the scattering plane, and hence the $a$-axis must be the spin axis of the the N\'{e}el order (See supplement). This $a$-axis G-type AFM structure breaks both the $b$-glide and $n$-glide [Fig.~\ref{figT}{\bf c}] that protects the nonsymmorphic semimetallicity \cite{yang2022quasi}.  

\subsection{Anomalous Hall effect and anomalous Nernst effect}
\begin{figure}[htbp]
\includegraphics[scale=0.6]{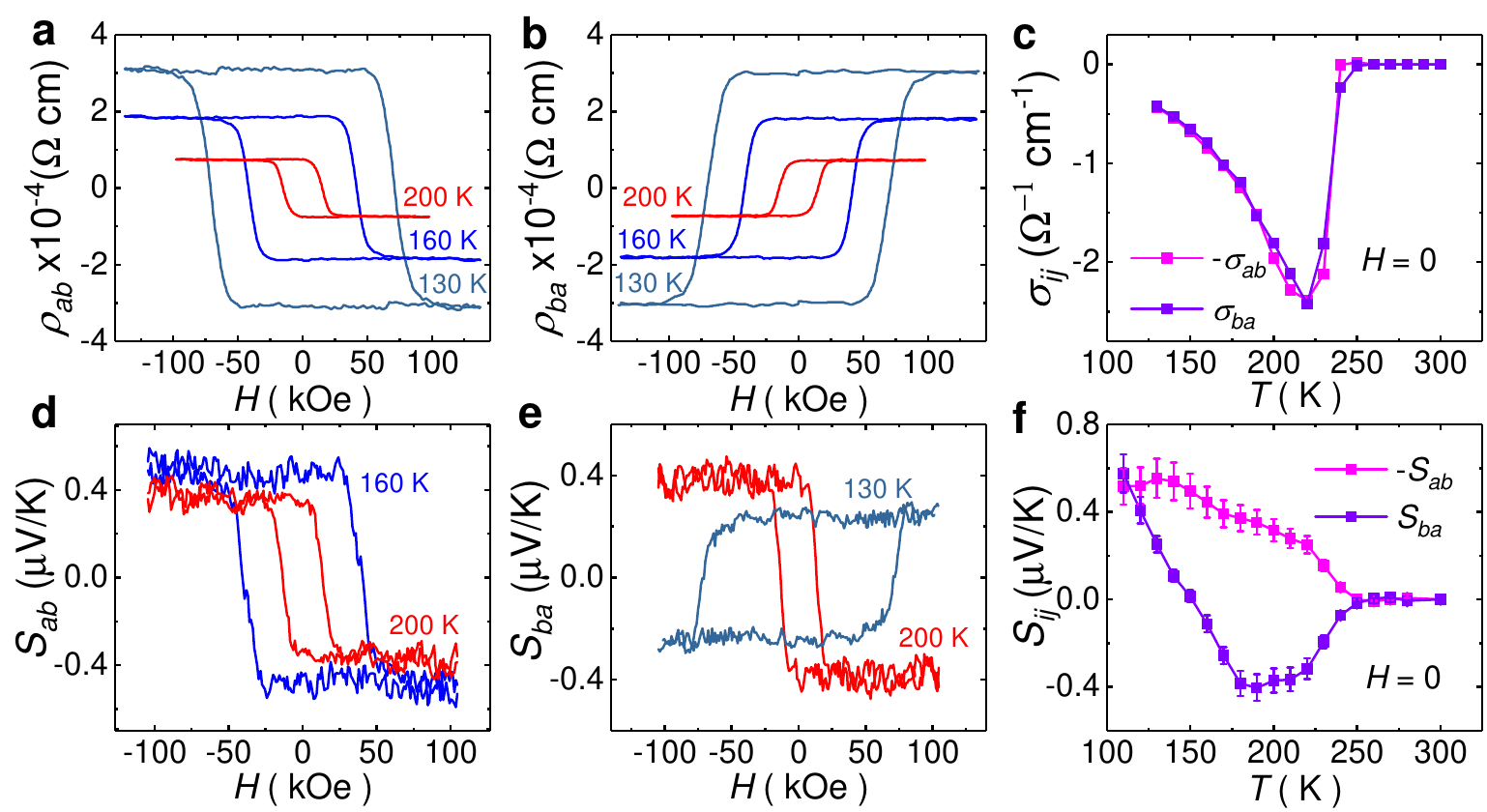}
\caption{\textbf {Anomalous Hall and Nernst effect of SrIr$_{0.8}$Sn$_{0.2}$O$_3$.} \textbf{a,b} Hysteresis loops of the anomalous Hall resistivity $\rho_{ab}$ and $\rho_{ba}$ at selected temperatures. The ordinary Hall resistivity is subtracted (see method). \textbf{c} Temperature dependence of spontaneous Hall conductivity. $\sigma_{ab}$ is multiplied by -1 to illustrate the antisymmetric relationship with $\sigma_{ba}$. \textbf{d,e} Hysteresis loops of the transverse thermopower $S_{ab}$ and $S_{ba}$ at selected temperatures.  \textbf{f} Temperature dependence of spontaneous Nernst thermopower. $S_{ab}$ is multiplied by -1 for comparison in a representation consistent with the Hall conductivity. The error bars represent the noise uncertainty.}\label{fig2}
\end{figure}

The corresponding magnetic point group of the G-type AFM order is $m'm'm$, which allows spontaneous Hall transport within the $ab$-plane.
As shown in Fig.~\ref{fig2}{\bf a}\&{\bf b}, well-defined hysteresis loops were indeed observed below the N\'{e}el transition in Hall resistivity $\rho_{ab}$ and $\rho_{ba}$ (See method) with the field $H//c$ which flips the $a$-axis N\'{e}el vector by flipping the canted moment along the $c$-axis [Fig.~\ref{figT}{\bf c}]. The rectangle-like shape of the hysteresis loop preserves the AHE down to zero field for a clear spontaneous signal. More importantly, the hysteresis loops of $\rho_{ab}$ and $\rho_{ba}$ mirror each other, well demonstrating their antisymmetric relation. Since the sample has a single orthorhombic domain, 
 the measured $\rho_{ab}$ and $\rho_{ba}$ correspond to two distinct elements of the resistivity tensor, and their antisymmetric relation shows that the Onsager reciprocal relation is complied in AHE. This relation is further demonstrated by the close matching between the spontaneous Hall conductivity -$\sigma_{ab}$ and $\sigma_{ba}$ as a function of temperature [Fig.~\ref{fig2}{\bf c}]. 

The observation of AHE suggests the occurrence of the ANE within the $ab$-plane which is indeed allowed under the $m'm'm$ magnetic point group. 
As shown in Fig.~\ref{fig2}{\bf d}\&{\bf e}, well-defined hysteresis loops were also observed in the ANE thermopower, $S_{ab}$ and $S_{ba}$, (See method) below the N\'{e}el transition with the field $H//c$. Since the TbScO$_3$ substrate is a paramagnetic insulator, the ANE hysteresis loops must originate from the SrIr$_{0.8}$Sn$_{0.2}$O$_3$ film. The coercivity and shape of the ANE loops are consistent with AHE at different temperatures. The spontaneous Nernst thermopower quickly exceeds the level of 0.1 $\mu$V/K within 10$\sim$15 K below the N\'{e}el transition, and goes above 0.5 $\mu$V/K at 100 K, which is comparable with the recently reported ferromagnetic and antiferromagnetic topological metals at zero magnetic field, such as Fe$_{3}$Ga (1 $\mu$V/K) \cite{sakai2020iron}, Mn$_{3}$Ge (1 $\mu$V/K) \cite{xu2020finite}, and Mn$_{3}$Sn (0.6 $\mu$V/K) \cite{ikhlas2017large}. More interestingly, there is significant asymmetry between $S_{ab}$ and $S_{ba}$. They are clearly not antisymmetric in sharp contrast to AHE. In fact, as seen in Fig.~\ref{fig2}{\bf f}, the spontaneous signals of $-S_{ab}$ and $S_{ba}$ even have opposite signs upon emerging below the N\'{e}el transition. While $-S_{ab}$ is positive and continues increasing with cooling, $S_{ba}$ makes a turn at around 200 K after reaching $-0.4$ $\mu$V/K and quickly increases by crossing the zero line onto the positive side. The sign change of $S_{ba}$ was confirmed by the inverted hysteresis loops at 200 K and 130 K shown in Fig.~\ref{fig2}{\bf e}. $S_{ab}$ and $S_{ba}$ are thus neither antisymmetric nor symmetric, indicating that the large ANE is driven by a nontrivial transverse transport mechanism that is absent in AHE.

\subsection{Asymmetric anomalous Nernst conductivity}
\begin{figure}[htbp]
\begin{center}
\includegraphics[scale=0.7]{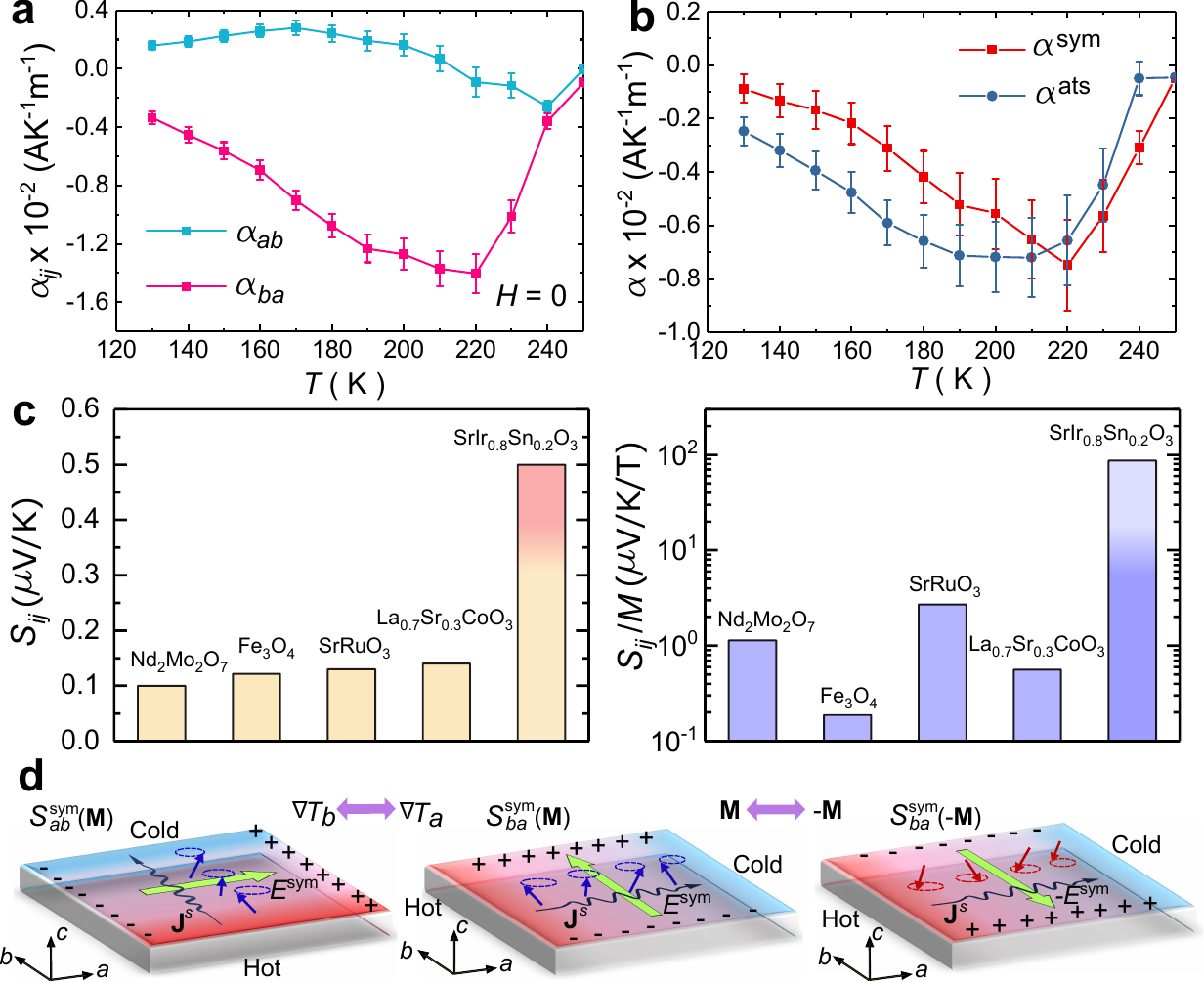}
\caption{\textbf{Asymmetric anomalous Nernst conductivity}. \textbf{a} Anomalous Nernst conductivity as a function of temperature at zero magnetic field. \textbf{b} Temperature dependence of symmetric and antisymmetric compoenents of the anomalous Nernst conductivity at zero magnetic field. \textbf{c} Comparison of $S_{ij}$ and $S_{ij}/M$ of SISO with other magnetic oxides with significant ANE, including Nd$_2$Mo$_2$O$_7$ \cite{hanasaki2008anomalous}, Fe$_3$O$_4$ \cite{ramos2014anomalous}, SrRuO$_3$ \cite{kan2019strain}, and La$_{0.7}$Sr$_{0.3}$CoO$_3$ \cite{soroka2021anomalous}. \textbf{d} The schematic of possible mechanism for the time-reversal odd and symmetric component of the ANE. The red and blue shade represent the hot and cold end, respectively. $\mathbf{J}^s$ represents a magnon-driven longitudinal spin current between the hot and cold ends. $E^{\mathrm{sym}}$ represent the transverse electric field induced by $\mathbf{J}^s$ via the inverse spin Hall effect. The left and middle configurations are related by exchanging the roles of the two axes, illustrating the symmetric relation. The middle and right configurations are related by reversing the magnetic order parameter $\textbf{M}$, illustrating the time-reversal odd nature of the mechanism. The error bars arise from propagation of uncertainty.}\label{fig3}
\end{center}
\end{figure}

To shed light onto the origin of the asymmetry between $S_{ab}$ and $S_{ba}$, it is necessary to extract the anomalous Nernst conductivity, $\alpha_{ab}$ and $\alpha_{ba}$. In an open-circuit setup at equilibrium, electric current $\mathbf{J}$ is zero, and temperature gradient $\nabla T$ produces both the longitudinal and transverse electric fields via $\mathbf{J}=\boldsymbol{\sigma} \cdot \mathbf{E}+\boldsymbol{\alpha} \cdot(-\nabla T)=0$, where $\boldsymbol{\sigma}$ and $\boldsymbol{\alpha}$ are the electrical conductivity and thermoelectric conductivity. Therefore, the transverse thermopower can be expressed as $S_{ba}=\rho_{bb}\left(\alpha_{ba}-S_{aa} \sigma_{ba}\right)$ and $S_{ab}=\rho_{aa}\left(\alpha_{ab}-S_{bb} \sigma_{ab}\right)$.
$S_{ab}$ and $S_{ba}$ may thus be asymmetric simply due to anisotropy of the longitudinal resistivity and thermopower. However, the longitudinal anisotropy here is too small to account for such large transverse asymmetry (Supplementary Fig.S4{\bf a}). After using the measured longitudinal resistivity and thermopower along both $a$- and $b$-axes (Supplementary Fig.S4{\bf b}\&{\bf c}), one can see the extracted spontaneous $\alpha_{ab}$ and $\alpha_{ba}$ in Fig.~\ref{fig3}{\bf a} are highly asymmetric. $|\alpha_{ba}|$ is one-order larger than $|\alpha_{ab}|$ in an extensive temperature range, which is what causes the transverse thermopower asymmetry. 

The asymmetry indicates that the anomalous Nernst conductivity has two components of different nature, i.e., a symmetric one $\alpha^{\mathrm{sym}}=(\alpha_{ba}+\alpha_{ab})/2$ and an antisymmetric one $\alpha^{\mathrm{ats}}=(\alpha_{ba}-\alpha_{ab})/2$. The latter likely shares the same origin as AHE through the Berry curvature of the electronic structure (See supplement). Its temperature dependence [Fig.~\ref{fig3}{\bf b}] is indeed similar to the AHE conductivity both with a non-monotonic behavior because both effects rely on thermally excited carriers across the charge gap \cite{fujioka2018charge} due to the insulating ground state. In other words, the effect is maximized after a sharp increase of the spontaneous time-reversal symmetry breaking field below the N\'{e}el transition and before the thermal excitation decrease takes over. $\alpha^{\mathrm{sym}}$ shows a similar initial increase followed by a slow decrease upon cooling as seen in Fig.~\ref{fig3}{\bf b}, indicating that both the magnetic ordering and the thermal excitation are necessary for the symmetric contribution as well. More importantly, $\alpha^{\mathrm{sym}}$ is comparable with $\alpha^{\mathrm{ats}}$ in size, which is the key in generating the large asymmetry between $\alpha_{ab}$ and $\alpha_{ba}$ and resulting in the large ANE thermopower. Without this asymmetry, both $S_{ab}$ and $S_{ba}$ would be significantly reduced, because the anomalous Hall conductivity also contributes to the transverse thermopower through Seebeck effect and it is compensating the anomalous Nernst conductivity contribution.

\section{Discussion}
The combination of the large ANE and the AFM order places SISO in a unique regime among transverse thermoelectric materials. We compared SISO with other magnetic oxides known to exhibit ANE. Unlike magnetic metals, both topology and magnetism in transition metal oxides originate from the $d$-electrons and thus naturally bear a strong mutual coupling as the exchange interaction an electron experiences when hopping from site to site could be on the scale of 10-100 meV \cite{jeong2023correlated}. AFM orders are common as well in Mott-Hubbard physics \cite{lee2006doping}. Figure~\ref{fig3}{\bf c} shows the comparison with Nd$_2$Mo$_2$O$_7$ \cite{hanasaki2008anomalous}, Fe$_3$O$_4$ \cite{ramos2014anomalous}, SrRuO$_3$ \cite{kan2019strain}, and La$_{0.7}$Sr$_{0.3}$CoO$_3$ \cite{soroka2021anomalous}, and clearly showcases the outstanding ANE of SISO. This superiority is particularly notable when considering the ANE thermopower normalized by the magnetization. Due to the AFM nature of SISO, the spontaneous magnetization arises from spin canting and is of a small magnitude \cite{cui2016slater}. Magnetization measurement is unfortunately not feasible here due to the strong paramagnetic signal from the TbScO$_3$ substrate. Therefore, we estimate the spontaneous magnetization ($\sim$0.035 $\mu_{\text{B}}$/f.u. at 150 K), based on values from both thin film and bulk samples \cite{cui2016slater,fujioka2018charge,yang2019epitaxial,negishi2019contrasted}. The normalized value exceeds the other magnetic oxides by at least one to two orders of magnitude. Although SISO may have a relatively smaller longitudinal electrical conductivity due to its insulating ground state, a large ANE is often much more important in Nernst applications, such as thermal radiation sensing and heat-flow meter \cite{mukherjee2022recent,zhou2020heat,mizuguchi2019energy,tanaka2023roll}. Apart from the consideration of energy-conversion efficiency, insulating thermoelectric materials may offer certain advantages in the context of integration with electronic circuits. For instance, the low electric and heat conductivities in the longitudinal direction may effectively reduce current noise and heat loss, which facilitates thermal management in low-power devices.
The existence of a correlated charge gap may also enable gating control \cite{guan2023ionic}, which is not possible in metals but could be important in integrating transverse thermoelectrics with semiconductor technologies.

Finally, we discuss the possible mechanism to generate the symmetric contribution that is time-reversal odd. And it must have a different origin from the time-reversal odd antisymmetric one. Given that spin-orbit coupling of the $J_{\mathrm{eff}}$ = 1/2 electrons is strong in SISO and the ANE vanishes above the N\'{e}el transition, a likely spin-mediated scenario is that the spin-motive force associated with the coherent propagation of thermally driven magnons drives a longitudinal spin current which then gives rise to a transverse electric voltage via the inverse spin Hall effect [Fig.~\ref{fig3}{\bf d}]~\cite{zelezny2017spin,he2021large,yang2023role}. A large spin Hall effect has indeed been reported in SrIrO$_3$ \cite{nan2019anisotropic,patri2018theory}. But the key of this mechanism is that the spin conductivity has a time-reversal odd part which is symmetric upon exchanging the two orthogonal directions (See supplement) because it comes from the quantum metric \cite{zelezny2017spin} and is irrelevant to the intrinsic Berry curvatures. This part, in combination with a time-reversal even spin current induced by the temperature gradient, gives rise to a time-reversal odd symmetric contribution to the transverse thermopower. Our observations thus shows that the transverse asymmetry can not only enhance the ANE thermopower but also reveal quantum geometry in a unique way thanks to the unconstrained matrix elements of the thermoelectric tensor. In principle, since the Onsager relation places no constraint, the transverse asymmetry may emerge as long as it is allowed by the point group symmetry, which in the case of SISO is the $m'm'm$ magnetic point group. Therefore, large asymmetry could be a rather general phenomena of magnetic topological materials.

In conclusion, we show that Sn-substituted SrIrO$_{3}$ has a large ANE asymmetry between the two transverse configurations due to the coexistence of significant symmetric and antisymmetric components. While the presence of the antisymmetric component can be well expected from the antisymmetric AHE, the symmetric one plays a key role in realizing the observed large ANE ($\sim$0.5 $\mu$V/K). This behavior highlights the fact that the Onsager reciprocal relation does not place symmetry constraint among the Nernst coefficients within the Seebeck tensor, and one could exploit both components to vanish and enhance the ANE. Although ANE materials usually exhibit AHE, the symmetric contribution may be left out or overlooked if assuming the same antisymmetric relation. Our work suggests that it may well exist in many materials as long as it is allowed under the magnetic point group. Separating the two components is thus critical in studying the ANE because they necessarily originate from different mechanisms. Our results suggest that SISO represents a class of ANE materials that are yet to be explored but have great potential for developing transverse thermoelectrics and Nernst applications with both contributions.

\section{Methods}
\subsection{Sample preparation}
SISO thin films were deposited from a polycrystalline target with a nominal SrIr$_{0.8}$Sn$_{0.2}$O$_{3}$ composition \cite{cui2016slater} onto (001)-oriented TbScO$_{3}$ single crystal substrates by using a pulsed laser deposition system (KrF excimer laser) equipped with an $in~situ$ reflection high-energy electron diffraction (RHEED) unit. Extra Ir was added when preparing the target to compensate potential loss of Ir during the deposition due to its low vapor pressure. During the deposition, the substrate temperature and the laser fluence were kept as 700$^o$ and 2 J/cm$^{2}$, respectively, with a constant oxygen pressure of 115 mTorr. The lattice constants extracted from synchrotron XRD measurements are $a$ = 5.723 \AA, $b$ = 5.454 \AA and $c$ = 7.961 \AA. The actual Sn concentration is likely different from the nominal one as shown in our previous work Ref.\cite{yang2019epitaxial} since Ir is expected to evaporate more than other elements during the deposition.

\subsection{X-ray diffraction and resonant X-ray magnetic scattering}
In-house X-ray diffraction (XRD) was performed using an X'Pert XRD Diffractometer with wavelength 1.54 \r{A}. Synchrotron XRD measurements were carried out at room temperature at Beamline 33-BM at the Advanced Photon Source of the Argonne National Laboratory and at Beamline ID4B at the Center for High-Energy X-ray Science of Cornell University to study the crystal structure of the thin films. Resonant X-ray magnetic scattering (RXMS) experiments were performed at Beamline 6-ID-B at the Advanced Photon Source of the Argonne National Laboratory. RXMS measurements were conducted at the Ir $L_3$-edge in the $\sigma-\pi$ channel under the vertical scattering geometry, and the reference azimuthal direction was deﬁned as the $b$-axis of the thin film. 

\subsection{Measurement of transport properties}
Transport measurements, including resistivity, Hall effect, and thermopower were measured in a Physical Property Measurement System (PPMS) with resistivity and thermal transport options. The thermopower measurements was calibrated with nickel and copper standard samples. The longitudinal measurements were conducted with the electric voltage probe being parallel to the applied electrical current and temperature gradient for resistivity and Seebeck effect, respectively. The transverse measurements were conducted with the electric voltage probe being normal to the applied electrical current and the applied thermal gradient for Hall effect and Nernst effect, respectively. For instance, when the electrical current or thermal gradient is applied along the $a$ direction, the electric voltage is measured along the $b$ direction for obtaining the Hall resistivity $\rho_{ba}$ or the Nernst thermopower $S_{ba}$, respectively. Note that $a$ and $b$ are crystallographic axes. The polarity of each quantity is defined by the positive direction of the chosen axis within a coordinate that was used consistently among all measurements.  Reversing the polarity of the voltage probe, for example, would redefine the obtained coefficients as $\rho_{-ba}$ and $S_{-ba}$. The anomalous Hall resistivity is extracted by subtracting the ordinary Hall component estimated from a linear fit to the high-field region.

To eliminate longitudinal leakage, the spontaneous Nernst thermopower was obtained by the formula $S_{ab}(B=0)= (S_{ab}(+0)-S_{ab}(-0))/2$, where the data at $B=+0$ and $B=-0$ was taken after opposite field sweeps. The samples were initially cooled from 300 K to 100 K under a magnetic field of $+2$T($-2$T). Then the field was reduced to $+0$T($-0$T) at 100 K, and transverse signals were measured as the sample was heated to various temperatures.

\subsection{DFT calculations}
The electronic band structures were calculated using density functional theory with the Full-Potential Local-Orbital program (FPLO) \cite{koepernik1999full}. Subsequently, symmetry-adapted tight-binding models were developed by projecting the Bloch wavefunction onto localized Wannier functions via automatic Wannierization workflow \cite{zhang2021different}. Hubbard correction is set to be $U$ = 1.3 eV, which gives rise to a local moment as 0.16 $\mu_B$ and global gap as 150 meV.

\section{Data availability}
All raw data generated in this study are provided in the Source Data file.

\bibliographystyle{naturemag}
\bibliography{refer.bib}

\section{Acknowledgements}
J. L. acknowledges support from the National Science Foundation under Grant No. DMR-1848269 and the Office of Naval Research (Grant No. N00014-20-1-2809). J. Y. acknowledges funding from the State of Tennessee and Tennessee Higher Education Commission (THEC) through their support of the Center for Materials Processing. S. Z. is supported by the Leibniz Association through the Leibniz Competition Project No. J200/2024. D.C. is supported by the U.S. Department of Energy under grant No. DE-SC0020254. L. H. acknowledges the support by the MGML infrastructure (project no. LM2023065). Use of the Advanced Photon Source, an Office of Science User Facility operated for the U.S. DOE, OS by Argonne National Laboratory, was supported by the U.S. DOE under Contract No. DE-AC02-06CH11357. This work is based on research conducted at the Center for High-Energy X-ray Sciences (CHEXS), which is supported by the National Science Foundation (BIO, ENG and MPS Directorates) under award DMR-1829070 and DMR-2342336. Y. Z. is supported by the Max Planck Partner lab for quantum materials from Max Planck Institute Chemical Physics of Solids.

\section{Author Contributions Statement}
J.L. conceived and directed the project. J.Y. and S.P. grew the thin films. J.Y., S.P., and D.G. characterized the samples. L.Horak. performed the in-house high-resolution x-ray diffraction measurement. D.G. designed the thermoelectric measurement. D.G., J.Y., and D.C. measured the electric and thermoelectric transport properties. S.Z. and Y.Z. performed the theoretical analysis. J.Y. and D.G. performed the synchrotron x-ray  experiments with the help of J.R., E.K., J.W.K., P.J.R., and L.Hao.. D.G., J.Y., Y.Z., and J.L. analyzed the results and wrote the manuscript with inputs from all other authors.

\section{Competing Interests Statement}
The authors declare no competing interests.

\end{document}